\documentclass[aps,prl,reprint,groupedaddress]{revtex4-1}

\usepackage{graphicx}
\usepackage{dcolumn}
\usepackage{bm}
\usepackage{subfigure}

\begin{document}
\bibliographystyle{plain}
\bibliographystyle{unsrtnat}


\title{Large valley splitting in monolayer WS$_2$ by proximity coupling to an insulating antiferromagnetic substrate}


\author{Lei Xu,$^{1,\dag}$ Ming Yang,$^2$ Lei Shen,$^3$ Jun Zhou,$^1$ Tao Zhu,$^1$ Yuan Ping Feng,$^{1,4,}$}
\email[$^\dag$Lei Xu: ]{xulei0553@gmail.com}
\email[$^*$ Yuan Ping Feng: ]{phyfyp@nus.edu.sg}
\affiliation{$^1$Department of Physics, National University of Singapore, Singapore 117542, Singapore}
\affiliation{$^2$Institute of Materials Research and Engineering, A*STAR, 2 Fusionopolis Way, Innovis, Singapore 138634, Singapore}
\affiliation{$^3$Department of Mechanical Engineering, National University of Singapore, Singapore 117575, Singapore}
\affiliation{$^4$Centre for Advanced 2D Materials and Graphene Research Centre, National University of Singapore, Singapore 117546, Singapore}


\date{\today}

\begin{abstract}
Lifting the valley degeneracy is an efficient way to achieve valley polarization for further valleytronics operations. In this work, we demonstrate that a large valley splitting can be obtained in monolayer transition metal dichalcogenides by magnetic proximity coupling to an insulating antiferromagnetic substrate. As an example, we perform first-principles calculations to investigate the electronic structures of monolayer WS$_2$ on the MnO(111) surface. Our calculation results suggest that a large valley splitting of 214 meV, which corresponds to a Zeeman magnetic field of 1516 T, is induced in the valence band of monolayer WS$_2$. The magnitude of valley splitting relies on the strength of interfacial orbital hybridization, and can be continually tuned by applying an external out-of-plane pressure and in-plane strain. More interestingly, we find that both spin and valley index will flip when the magnetic ordering of MnO is reversed. Besides, owing to the sizeable Berry curvature and time-reversal symmetry breaking in the WS$_2$/MnO heterostructure, a spin and valley polarized anomalous Hall current can be generated in the presence of an in-plane electric field, which allow one to detect valleys by the electrical approach. Our results shed light on the realization of valleytronic devices using the antiferromagnetic insulator as the substrate.
\end{abstract}


\maketitle

\section{I. Introduction}
\indent The utilization of valley degree of freedom, which is also called valley pseudospin, as the information carrier is the main context of valleytronics \cite{1Xiao_RMP,2Xiao_PRL,3Schaibley_RNM}. Many systems, such as graphene, silicene, bismuth thin film and AlAs quantum wells, have been studied to generate, detect and control the valley pseudospin \cite{3Schaibley_RNM,4Rycerz_NP,5Ezawa_PRB,6Zhu_NP}. Valleys, which label the degenerate energy extreme of conduction band or valence band at some special $k$ points, have large separation in the momentum space which enables valley pseudospin very robust against phonon and impurity scattering. Once the structural inversion symmetry is broken, the carriers at these inequivalent valleys are associated with some valley-contrasting physical quantities, like Berry curvature $\Omega$ and orbital magnetic moment m \cite{2Xiao_PRL,7Feng_PRB}. These distincitve properties make the generation and manipulation of valley pseudospin accessible by means of electric, magnetic and optical ways \cite{2Xiao_PRL,9Yao_PRB,8Zeng_NN,13Cao_NC,14Mak_NN,10Aivazian_NP,11MacNeill_PRL,15Li_PRL}.

\indent Monolayer transition metal dichalcogenides (TMDs) MX$_2$ (M=Mo, W; X=S, Se, Te) are a class of two dimensional (2D) materials with direct band gaps, where both conduction band and valence band edges are located at the corners of 2D Brillouin zone. Two inequivalent valleys are formed at K and K$^{'}$ points as a result of C$_{3v}$ crystal symmetry of pristine monolayer TMDs, and they constitute a binary index for low energy carriers \cite{7Feng_PRB,30xiao_PRL}. Due to the strong spin-orbit coupling (SOC) and intrinsic inversion symmetry breaking, monolayer TMDs are considered as good candidates for valleytronic applications \cite{3Schaibley_RNM,29liu_CSR,12Xu_NP}. The realization of valley polarization, which breaks the balance of carriers in the inequivalent valleys, is an indispensable step for further manipulation of valley pseudospin. However, the number of carriers at K and K$^{'}$ valleys is same as required by the time reversal symmetry.\\
\indent Since the orbital magnetic moments in the two valleys are opposite, optical pumping with circular polarized light has been both theoretically and experimentally demonstrated to be able to achieve valley polarization \cite{8Zeng_NN,13Cao_NC,14Mak_NN}. Nevertheless, as a dynamics process, optical pumping is difficult to manipulate robustly and not applicable for practical valleytronic applications. Interestingly, the valley index and spin index are locked to each other, making it possible to coherently control these two degrees of freedoms. Therefore, the magnetic field can be applied to lift the valley degeneracy, where spin polarization is accompanied by a valley polarization. Indeed, valley splitting induced by an external magnetic field has been observed in the experiment, whereas the efficiency is very low as 1T magnetic field can only give rise to a splitting of 0.1-0.2 meV\cite{10Aivazian_NP,11MacNeill_PRL,15Li_PRL}. On the other hand, some theoretical works reported that doping with transition metal atoms can establish a considerable intrinsic magnetic field in monolayer TMDs, and a large permanent valley polarization will be generated\cite{40singh_AM,41cheng_PRB}. However, these metallic atoms tend to form a cluster and can significant increase the scattering rate, thus are detriment to the device performance. Moreover, it was predicted that a valley splitting over 300 meV can be generated in monolayer MoTe$_2$ by magnetic proximity coupling to a ferromagnetic insulator EuO\cite{16Zhang_AM,17Qi_PRB}. And recent experiment also found an enhanced valley splitting in monolayer WSe$_2$ when deposited on the EuS substrate\cite{18Zhao_NN}.\\
\indent In fact, the insulating ferromagnetic materials are very rare and always have a low Curie temperature, but antiferromagnetic insulators are common in nature and easy to be obtained. Hence, it's important and timely to know whether an antiferromagnetic insulator can also induce a valley splitting in monolayer TMDs through magnetic proximity interaction. In this work, we try to answer this question by using first-principles calculations to study valley splitting of WS$_2$ monolayer on the antiferromagnetic MnO (111) substrate, due to their small lattice mismatch and the considerable SOC effect in monolayer WS$_2$. Our calculation results show that the valence band of monolayer WS${_2}$ is well preserved and almost free of hybridization with the substrate, but the conduction band has a strong hybridization with the substrate. Due to the magnetic proximity effect, the valley degeneracy has been lifted and a sizeable valley splitting of 214 meV has been observed in the valence band of monolayer WS$_2$. The magnitude of splitting can be continually tuned by applying an external pressure and strain. A finite and fully spin and valley polarized anomalous Hall conductivity can be obtained  when the Fermi level lies between two valley extrema, which makes WS$_2$/MnO heterostructure very appealing for both spintronic and valleytronic applications.\\

\section{II. Computational details}
\indent Our first-principles calculations are performed by using Vienna $ab$ $initio$ simulation package (VASP)\cite{21Kresse_PRB} with generalized gradient approximation (GGA) of Perdew-Burke-Ernzerhof (PBE) functional\cite{22Perdew_PRL}. The ion-electron interaction is treated by projector augmented wave (PAW) method\cite{23Bl_PRB}, and the van der Waals interaction is taken into consideration using DFT-D3 method\cite{24Grimme_JCP}. Electron wave function is expanded on a plane wave basis set with a cut off energy of 500 eV. 12$\times$12$\times$1 $\Gamma$-centred Monkhorst-Pack grids are adopted for Brillouin-zone integration. A vacuum slab more than 20 \textup{\AA} is applied along the z direction(normal to the interface) to avoid spurious interaction between repeated slabs. Structural relaxation is carried out using the conjugate-gradient algorithm until the total energy converges to $10^{-5}$  eV and the Hellmann-Feynman force on each atom is less than 0.01 eV/\textup{\AA}, respectively. In order to include the strong on-site Coulomb interaction in MnO, the Coulomb and exchange parameters, $U$ and $J$, are set to 6.9 eV and 0.86 eV for $d$ orbital of Mn atom, respectively \cite{25Anisimov_PRB}.\\
\indent For the calculation of Berry curvature and anomalous Hall conductivity of monolayer WS$_2$ on the MnO substrate, we employ maximally localized Wannier function method\cite{34marzari_RMP} as implemented in the WANNIER90 package\cite{35mostofi_CPC}. Ten $d$ orbitals of W atom and six $p$ orbitals of each S atom are selected as the initial orbital projections, and a finer 27$\times$27$\times$1 uniform k-grid is used for the construction of maximally localized Wannier function. The spread of total Wannier functions can converge to $10^{-10} \textup{\AA}^2$ within 2000 iterative steps.

\section{III. Results and Discussion}
\indent Below the Neel temperature of $T_N=118$K, the bulk MnO adopts a rock-salt structure but with rhombohedrally distorted B1 symmetry\cite{20Shaked_PRB}. The optimized lattice constant of bulk MnO is 4.53 \textup{\AA}, which is close to the experiment measurement value of 4.44 \textup{\AA}\cite{26wyckoff}. Besides, our DFT+U calculations also predict that bulk MnO is a type-II antiferromagnetic insulator with a band gap of 2.1 eV and the magnetic ordering along (111) direction. The calculated magnetic moment on each Mn atom is 4.66 $\mu_B$, which is in good agreement with the experimental result of 4.58 $\mu_B$ \cite{27Fender_JCP}. Based on the optimized structures, the MnO (111) slab and monolayer WS$_2$ has an in-plane lattice constant of 3.203 \textup{\AA} and 3.184 \textup{\AA}, respectively, with a lattice mismatch of 0.6$\%$. We fix the in-plane lattice constant of WS$_2$/MnO heterostructure to the value of MnO(111) slab, thus a small tensile strain is applied in the monolayer WS$_2$. The substrate is constructed by six bilayers of MnO, and the bottom layer is always terminated by O atoms which are passivated with hydrogen atoms to avoid surface states.

\begin{figure}
\includegraphics[width=0.5\textwidth]{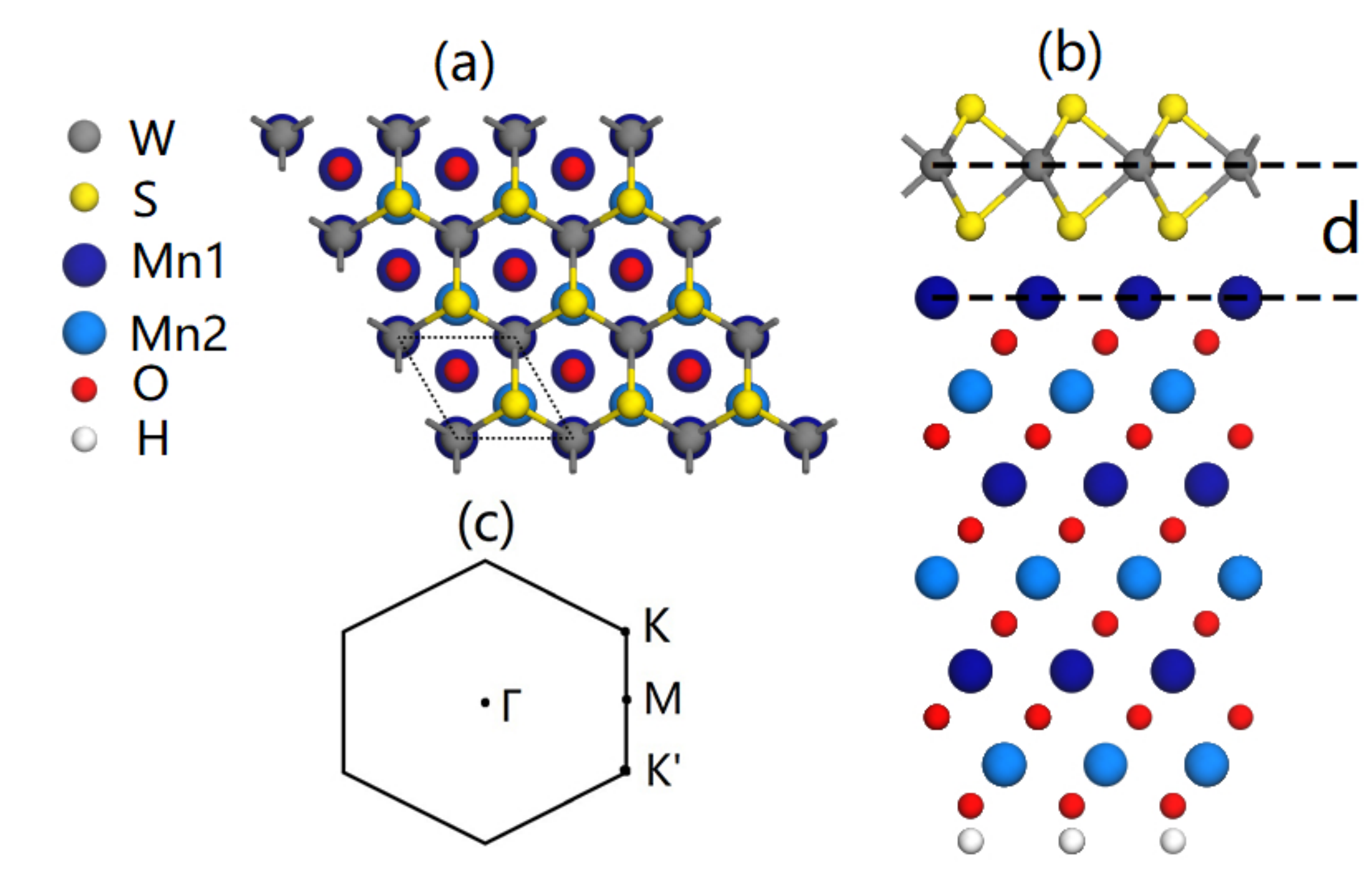}
\caption{\label{fig1} (a) Top view and (b) side view of the WS$_2$/MnO heterstructure, Mn1 and Mn2 represent the Mn atoms with opposite spin polarization. The defined interface distance is denoted by d. (c) First Brillouin zone of the WS$_2$/MnO structure with high symmetry points, and the primitive unit cell is shown by dotted lines in (a).}
\end{figure}

\indent There are two possible top surface terminations of the MnO(111) substrate, one is terminated by Mn atoms and the other is terminated by O atoms. The binding energy, which is defined as the energy difference between WS$_2$/MnO heterostructure and isolated systems, for the Mn-terminated top surface is calculated to be 0.33 eV larger than that of the O-terminated top surface. It indicates that Mn-terminated substrate has a much stronger interaction with monolayer WS$_2$. Besides, monolayer WS$_2$ is much closer to the magnetic Mn atoms in Mn-terminated case, so the magnetic proximity effect is expected to be more significant. Therefore, we will focus on monolayer WS$_2$ on the Mn-terminated MnO(111) surface in the following discussion. For the interfacial configurations, we have investigated six possible constructions by considering high symmetrical positions, namely, the topmost Mn atom or O atom in the substrate directly below W atom, S atom or hexagonal hollow site of monolayer WS$_2$. Among these six configurations, the one with topmost Mn atom directly below the W atom and topmost O atom sits below the hexagonal hollow site is the most stable one, as shown in Fig. 1. The separation between surface Mn atoms and W atoms, which is defined as interfacial distance d, is calculated to be 3.645 $\textup{\AA}$, and such a small distance implies that the MnO substrate would cause important impacts on the monolayer WS$_2$. Indeed, we find a magnetic moment of 0.05 $\mu_B$ has been induced on W atom, at the same time, the top and bottom S atoms acquire a magnetic moment of 0.01 and 0.02 $\mu_B$, respectively. It is noted that these magnetic moments are all ferromagnetic coupled with the interfacial Mn(Mn1) atom. The induced magnetism will break the time reversal symmetry of monolayer WS$_2$, and we expect the valley degeneracy should be lifted simultaneously.

\begin{figure}
\includegraphics[width=0.5\textwidth]{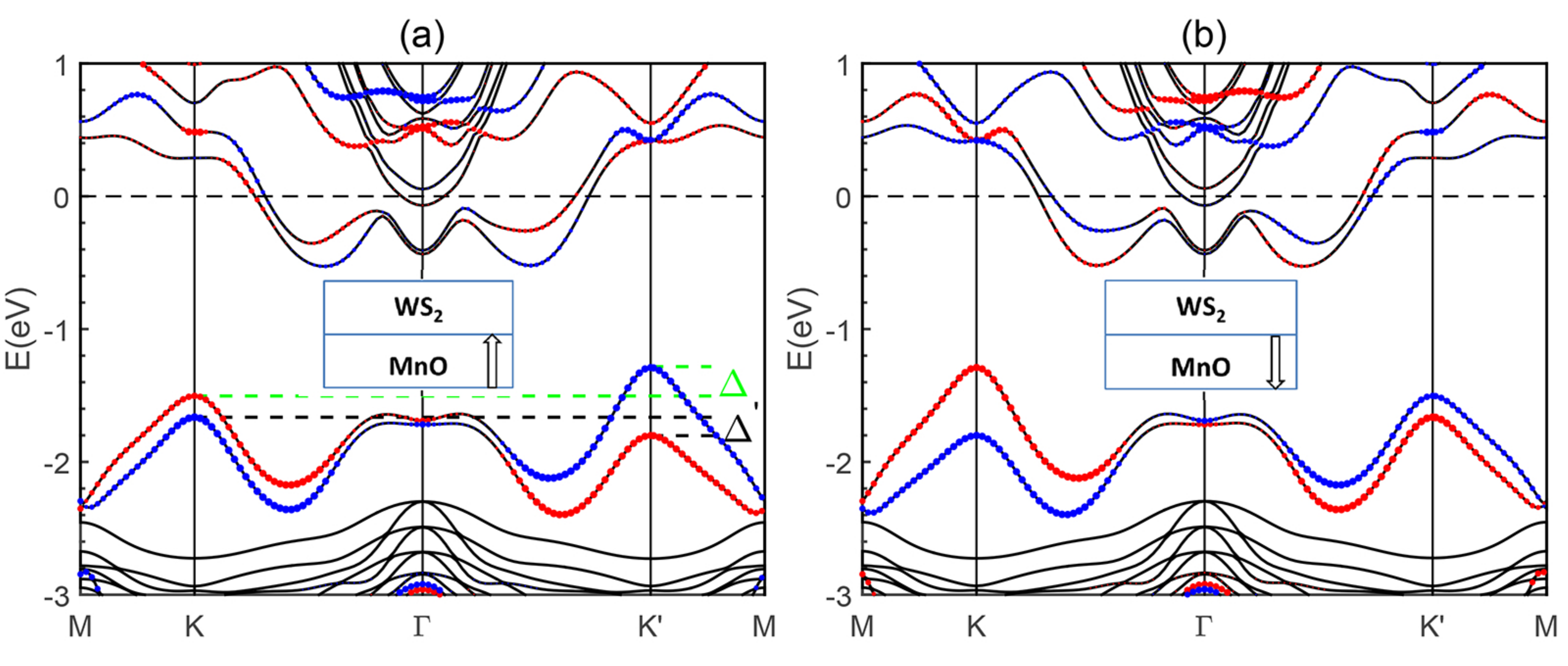}
\caption{\label{fig2} (a)-(b) Band structure of WS$_2$/MnO heterostructure with SOC for surface Mn(Mn1) atoms magnetized upward and downward, respectively, the spin projections for monolayer WS$_2$ states along positive (spin up) and negative (spin down) z axis are denoted by red and blue weighted solid circles, respectively. The magnitude of valley splitting in the first and second valence band are denoted by $\Delta$ and $\Delta^{'}$ in (a). The empty arrow in the inset shows the directions of surface Mn(Mn1) atoms' magnetic ordering.
}
\end{figure}

\indent In Fig. 2a, we show the band structure of WS$_2$/MnO heterostructure with SOC and surface Mn(Mn1) atoms magnetized upward. The spin projections for monolayer WS$_2$ states along positive (spin up) and negative (spin down) z directions are indicated by red and blue weighted solid circles, respectively. As can be seen, the conduction band of monolayer WS$_2$ has a strong orbital hybridization with the substrate, and its valley characteristic has been partly destroyed. Based on Bader analysis\cite{28henkelman_CMS}, we find that an amount of 0.49 $e^-$ is transferred from substrate to monolayer WS$_2$. Consequently, the Fermi level is shifted into the conduction band. In contrast, the valence band of monolayer WS$_2$ is little affected by the substrate and the two valleys (K and K$^{'}$) are well preserved. The spin polarized density of states (SOC not included), as shown in Fig 3, also illustrates that the conduction band edge states are contributed by both WS$_2$ and MnO, whereas the valence band edge states are only originated from WS$_2$ orbitals. In addition, we note that there is an energy shift of around 177 meV between the valence band maxima of the two spin channels, which is resulted from the substrate induced magnetism in monolayer WS$_2$.

\begin{figure}
\includegraphics[width=0.5\textwidth]{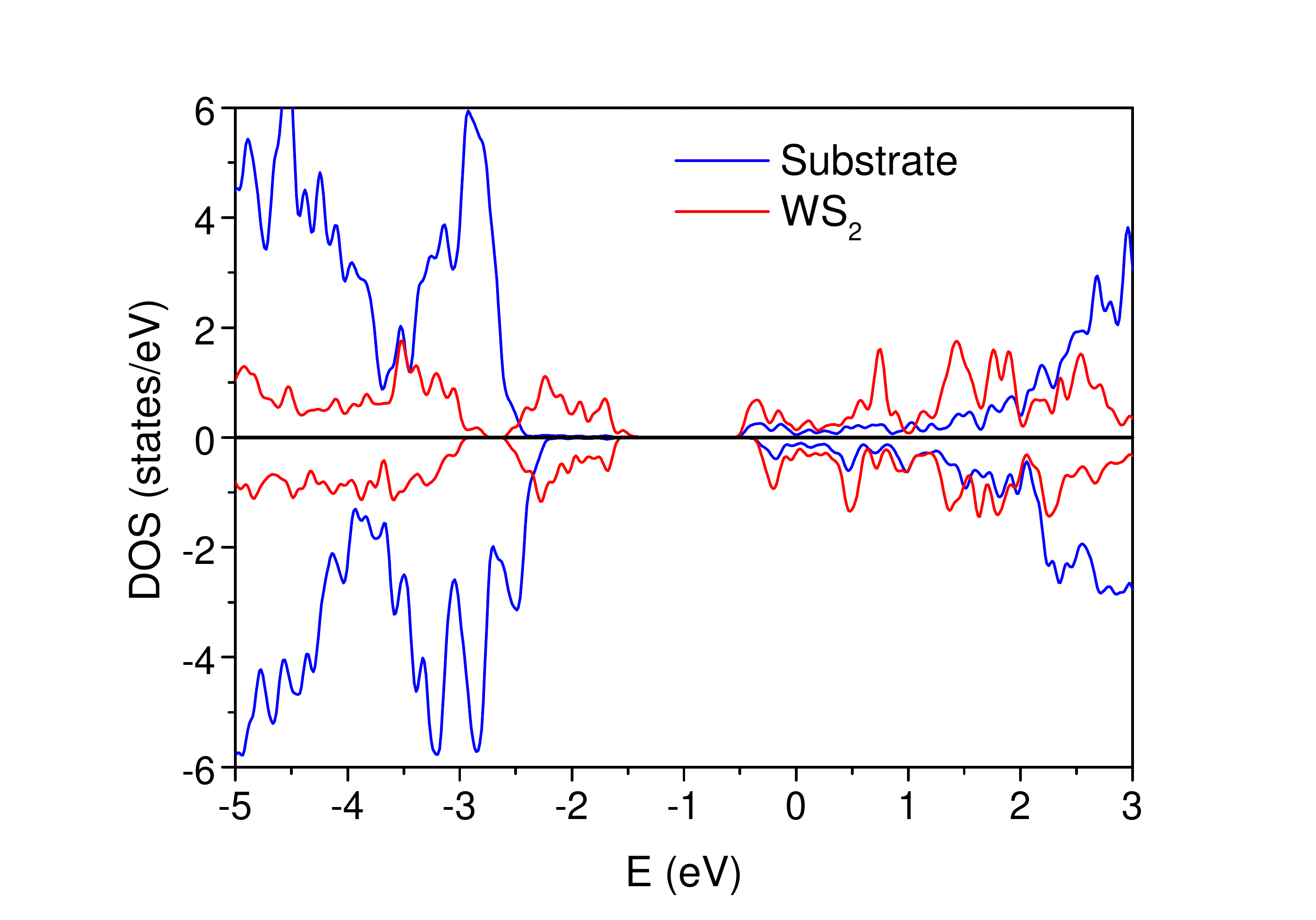}
\caption{\label{fig3} Spin polarized partial density of states of the WS$_2$/MnO heterostructure, the Fermi level is set to 0.
}
\end{figure}

\indent For freestanding monolayer WS$_2$, there is a large spin splitting in both K and K$^{'}$ valleys due to the inversion symmetry breaking and strong SOC, while the spin up band in one valley is energy degenerate with spin down band in the other valley as a result of time reversal symmetry\cite{7Feng_PRB}. However, as shown in Fig 2a, the valley degeneracy in monolayer WS$_2$ has been lifted when placed onto MnO substrate. Here, we only consider the valleys in the valence band of monolayer WS$_2$, and we define the valley splitting $\Delta$ as the energy difference between the two valley extrema, as denoted in Fig. 2a. Within this definition, the valley splitting in the valence band of monolayer WS$_2$ is found to be 214 meV, which is sizeable and comparable to the valley splittings predicted from MoTe$_2$/EuO and MoS$_2$/EuS heterstructures\cite{16Zhang_AM,17Qi_PRB,38liang_NS}.

\begin{figure}
\includegraphics[width=0.5\textwidth]{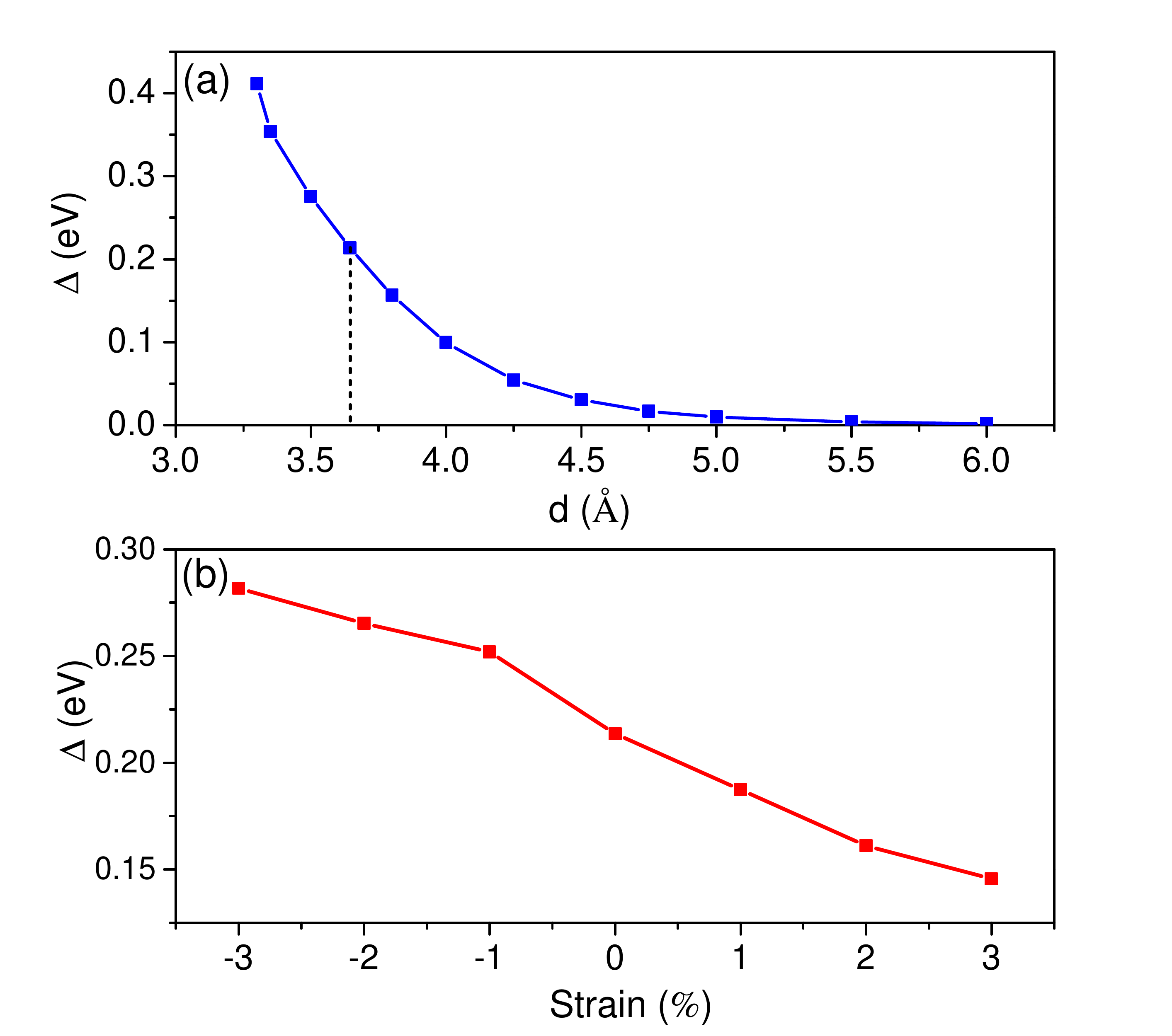}
\caption{\label{fig4} Valley splitting $\Delta$ as a function of (a) the interfacial distance d, where the equilibrium distance is denoted by dashed line, and (b) in-plane strain.
}
\end{figure}

\indent The magnetic proximity effect induced valley splitting is mediated with the interfacial orbital hybridization between monolayer WS$_2$ and MnO, and we expect its magnitude can be modulated by changing the hybridization strength. In the experiment, one can apply an external perpendicular pressure or insert a buffer layer to the heterostructure to adjust the interface distance, and then the interfacial orbital hybridization will be altered\cite{36zhao_NC,31nayak_NC,37farmanbar_PRB}. Fig. 4a shows the variation of valley splitting as a function of interface distance. It can be seen that the splitting is very sensitive to the separation, for example, the valley splitting can reach a value over 0.4 eV for a slightly smaller distance of 3.3 $\textup{\AA}$, but almost vanishes when the separation is larger than 5.5 $\textup{\AA}$, which implies that the magnetic proximity coupling is a short-range effect. On the other hand, strain effect also has an important influence on the band hybridization. By calculating the band structures of WS$_2$/MnO heterostructure under different in-plane strains, we find that the valley splitting also has a strong dependence on the applied external strain. As shown in Fig. 4b, a compressive strain can increase the valley splitting due to the enhanced hybridization, while a tensile strain will decrease the valley splitting. Hence, we can continually tune the magnitude of valley splitting by external pressure and strain methods.

\indent In order to have a better understanding of the magnetic proximity interaction induced large valley splitting, we construct a low-energy effective Hamiltonian based on the k.p model \cite{30xiao_PRL}. The Hamiltonian is expressed as:
\begin{eqnarray}
H=at(\tau{k_x}{\hat{\sigma}_x}+{k_y}{\hat{\sigma}_y})+\frac{\Delta}{2}{\hat{\sigma}_z}-\lambda{\tau}\frac{{\hat{\sigma}_z}-1}{2}{\hat{s}_z}\nonumber\\
+\frac{{\hat{\sigma}_z}-1}{2}({\hat{s}_z}+\tau\alpha){B}
\end{eqnarray}
where $a$, $t$, $\Delta$, $2\lambda$, $\alpha$, and $B$ are the lattice constant, effective hopping parameter, band gap, SOC strength, orbital magnetic moment and effective Zeeman magnetic field, respectively. $\hat\sigma$ is the Pauli spin matrix which is constructed on the basis $|d_{z^2}\rangle$ and $\frac{1}{\sqrt{2}}|d_{x^2-y^2}+i\tau{d_{xy}}\rangle$. Besides, $\tau=\pm{1}$ and ${\hat{s}_z}=\pm{1}$ are the spin index and valley index, respectively. The first three terms of the Hamiltonian describe the low energy band dispersion of pristine monolayer WS$_2$, while the last term accounts for the proximity induced exchange energy. It should be emphasized that the spin and valley degeneracy is still remained in the conduction band but lifted in the valence band for this Hamiltonian.

\indent It can be found that the presence of antiferromagnetic substrate MnO introduces a Zeeman magnetic field $B$, which lifts the valley degeneracy by coupling to the orbital and spin magnetic moment in the monolayer WS$_2$. Based on the eigenvalues of the Hamiltonian, we can deduce that the valley splitting $\Delta$ is $2(1+{\alpha})B$. To determine the value of $B$, we introduce another valley splitting $\Delta^{'}=2(1-{\alpha})B$, which is the splitting of second valence band, as shown in Fig. 2a. By fitting the two valley splittings with the first-principles calculation results($\Delta$=214 meV and $\Delta^{'}$=137 meV), an orbital magnetic moment $\alpha$ of 0.22 \cite{39note} and an effective Zeeman field $B$ of 87.75 meV can be obtained. The substrate induced effective Zeeman field corresponds to an equivalent magnetic field of 1516 T, which indicates a huge perpendicular magnetic field is built into the monolayer WS$_2$ through the magnetic proximity interaction.

\indent When we tune the Fermi level to the energy window between two valley extrema by a gate voltage or hole doping the WS$_2$/MnO heterostructure, both spin and valley polarization will be achieved. More interestingly, we find that both spin and valley index of the transport carriers can be flipped when reversing the magnetic ordering of MnO, as shown in Figs. 2a-2b. In experiment, one can deposit MnO onto a hard magnet material (like FePt or CoPt), which can not only pin the MnO's magnetic ordering, but also help to flip it in the presence of an external magnetic filed.

\begin{figure}
\includegraphics[width=0.5\textwidth]{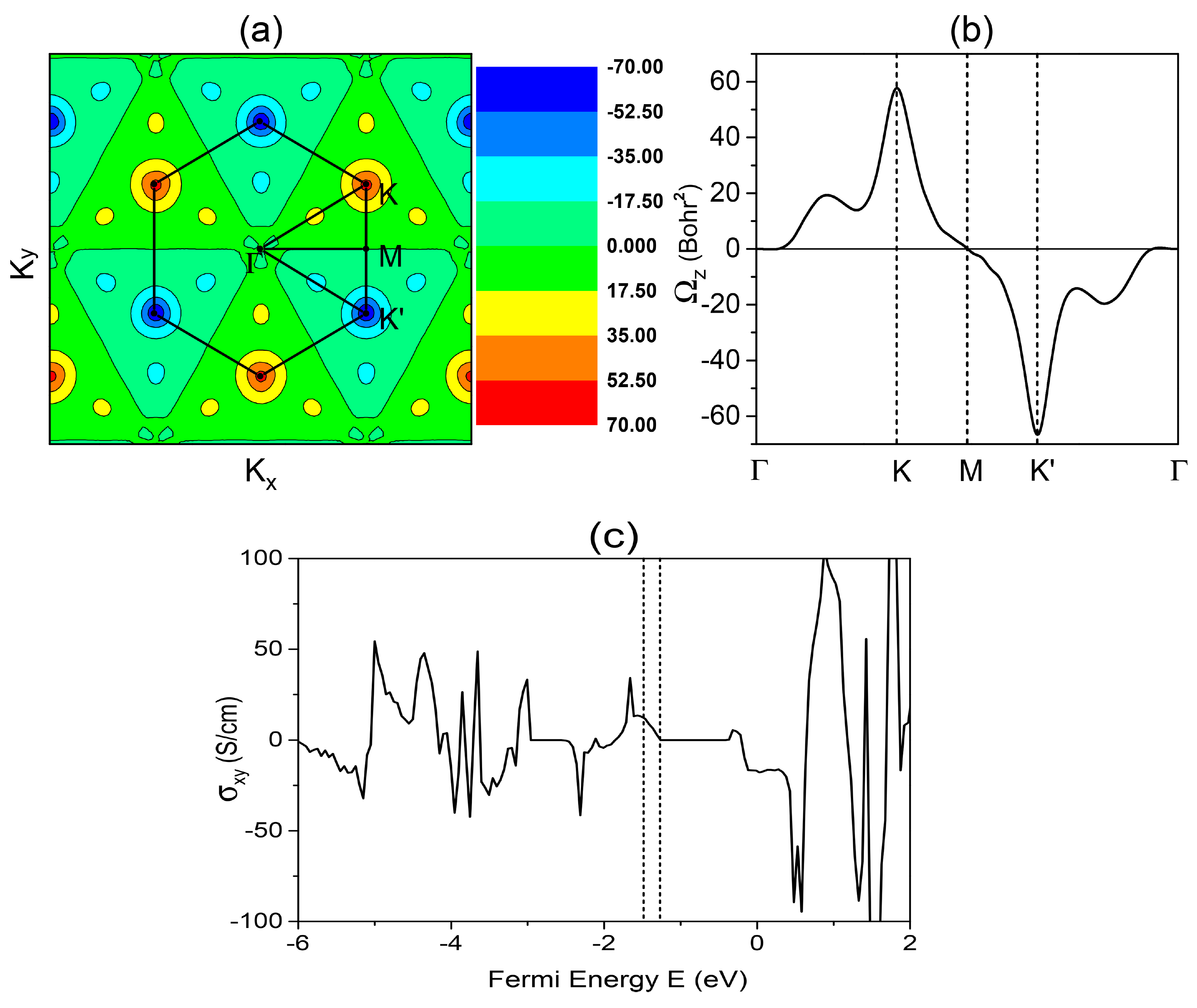}
\caption{\label{fig5} Calculated Berry curvature of monolayer WS$_2$ on the MnO substrate (a) over 2D Brillouin zone and (b) along high symmetry lines. (c) The calculated intrinsic anomalous Hall conductivity $\sigma_{xy}$ as a function of Fermi energy, the two dashed lines denote the two valley extrema.
}
\end{figure}

\indent Due to the intrinsic inversion symmetry breaking in monolayer WS$_2$, the charge carriers in the K and K$^{'}$ valleys will acquire a nonzero Berry curvature $\Omega$ along the out of plane direction (z axis). As derived from the Kudo formula \cite{32yao_PRL,33thouless_PRL}, the Berry curvature can be written as a summation of all occupied contributions:
\begin{eqnarray}
\Omega(k)=-\sum_n\sum_{n\ne{n^{'}}}{f_n}\frac{2\mathrm{Im}\langle{\psi_{nk}|v_x|\psi_{n^{'}k}}\rangle\langle{\psi_{n^{'}k}|v_y|\psi_{nk}}\rangle}{(E_n-E_{n^{'}})^2}
\end{eqnarray}
where $f_n$ is Fermi-Dirac distribution function, and $v_{x(y)}$ is the velocity operator.  $|\psi_{nk}\rangle$ is the Bloch wave function with eigenvalue $E_n$. In Figs. 5a and 5b, we show the calculated Berry curvature in the 2D Brillouin zone and along high symmetry lines, where the Fermi level has already been shifted into the WS$_2$ band gap. Obviously, the Berry curvature is sizeable and takes opposite signs in the vicinity of K and K$^{'}$ valleys, which reveal that the valley-contrasting characteristic is still remained in monolayer WS$_2$ even strongly hybridized with the MnO substrate. Under an in-plane  longitudinal electric field, the Berry curvature will give rise to an anomalous transverse velocity $v_\perp$  for Bloch electrons, $v_\perp\sim{E}\times{\Omega}(k)$ \cite{1Xiao_RMP}. Thus, charge carriers in the K and K$^{'}$ valleys will move in opposite directions due to the valley-contrasting Berry curvature.

\begin{figure}
\includegraphics[width=0.5\textwidth]{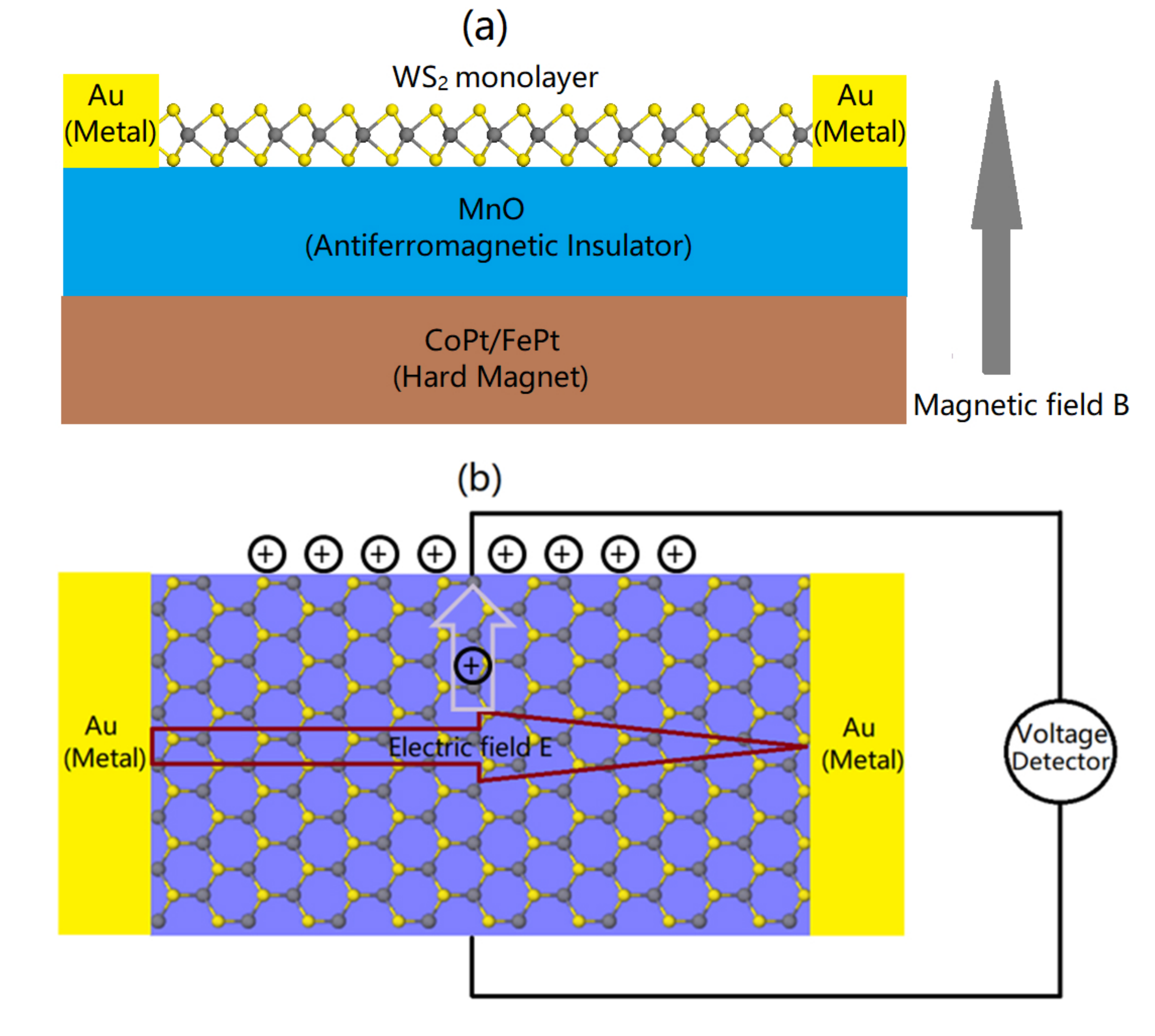}
\caption{\label{fig6} (a) Top view and (b) side view of proposed valleytronic device.
}
\end{figure}

\indent Owing to the giant valley splitting and sizeable Berry curvature in the monolayer WS$_2$, an anomalous Hall current would be observed when an in-plane electric field is applied to the WS$_2$/MnO heterostructure. By integrating the Berry curvature over the 2D Brillouin zone, we can obtain the intrinsic anomalous Hall conductivity $\sigma_{xy}$\cite{1Xiao_RMP,32yao_PRL}. In Fig. 5c, we show the calculated $\sigma_{xy}$ as a function of Fermi energy. As can be seen, when the Fermi level lies between the valence band maxima of K and K$^{'}$ valleys, as denoted by the dashed lines, a fully spin and valley polarized Hall conductivity will be generated. This provides us a method to detect the valley pseudospin by electric measurement and forms the basis for the application of valleytronic device. In Fig. 6, we propose a device to experimentally investigate the valley anomalous Hall effect, where antiferromagnetic insulator is not only severed as a substrate but also to achieve valley polarization in its supported monolayer WS$_2$. Besides, a hard magnet is used to pin the magnetic ordering of antiferromagnetic insulator. When hole doping the system to enable the Fermi level lies between the two valleys, the transport carriers, i.e. spin down holes from K$^{'}$ valley(Fig. 2a), will move toward upside in the presence of an in-plane external electric field due to their negative Berry curvatures, as shown in Fig. 6b. The accumulated holes will result in a net measurable voltage along the transversal direction, which can be experimentally detected by a voltmeter. Once the magnetic ordering of MnO substrate is reversed by an external magnetic field, spin up holes from K valley(Fig. 2b) will act as free carriers and move toward the downside since they have positive Berry curvatures, and then a voltage with opposite sign will be detected. It should be noted that the transport carriers have particular polarity for charge, spin and valley, and the anomalous Hall current is a combination of all three of them.

\section{IV. Conclusion}
In summary, through magnetic proximity coupling to the insulating antiferromagnetic MnO substrate, a large valley splitting of 214 meV is induced in the valence band of monolayer WS$_2$ based on our first-principles calculations. The magnetic proximity interaction is mediated with interfacial orbital hybridization, as a result the induced valley splitting shows a strong dependence on the interface distance and strain. Besides, the sizeable and valley-contrasting Berry curvature still remained in monolayer WS$_2$ despite its strong interaction with the MnO substrate. Due to the large valley splitting and time reversal symmetry broken in WS$_2$/MnO heterostructure, a finite and fully spin and valley polarized anomalous Hall conductivity can be obtained when the Fermi level is shifted between the maxima of two valleys. Therefore, WS$_2$/MnO heterostructure provides a good platform to detect the valley pseudospin and study the anomalous Hall effect. These findings are also expected applicable to other valley materials coupling to insulating antiferromagnetic substrates.\\

\section{V. Acknowledgement}
The authors acknowledge computational resources provided by the Centre for Advanced 2D Materials (CA2DM) at the National University of Singapore. M. Y thanks funding support  from Singapore A*STAR 2D PHAROS project: 2D devices $\&$ materials for ubiquitous electronic,
sensor and optoelectronic applications (Project No: SERC 1527000012).

\subsection{}
\subsubsection{}

\bibliography{apssamp}

\begin{thebibliography}{40}
\bibitem[39n()]{39note}
Here, in order to keep same with ${\hat{s}_z}$, $\alpha$ is unitless and we
  place its unit $\mu_{B}$ into {Z}eeman field ${B}$.
\end{thebibliography}

\end{document}